\documentclass[aps,pra,showpacs,twoside,twocolumn,10pt]{revtex4-1}
\usepackage[colorlinks=true, citecolor=red, urlcolor=blue ]{hyperref}
\usepackage{wrapfig,graphicx}

\usepackage{epsfig,newlfont,amssymb,amsfonts,amsmath,bm,subfigure,palatino,mathtools,amsthm,braket,times,soul,enumitem,color}
\usepackage{float}
\usepackage{framed}
\usepackage[normalem]{ulem}
\newcommand{\stkout}[1]{\ifmmode\text{\sout{\ensuremath{#1}}}\else\sout{#1}\fi}
\usepackage[utf8]{inputenc}
\usepackage{array}
\usepackage{graphics}

\newcommand{\ketbra}[2]{|#1\rangle \langle #2|}

\begin{document}


\title{Optimal quantum resource generation in coupled transmons immersed in Markovian baths}

\author{Tanaya Ray, Ahana Ghoshal, Debraj Rakshit, and Ujjwal Sen}
\affiliation{Harish-Chandra Research Institute,  A CI of Homi Bhabha National Institute, Chhatnag Road, Jhunsi, Prayagraj 211 019, India}

\begin{abstract}

We analyze the quantum resource generation of capacitively-coupled multilevel transmon circuits surrounded by bosonic baths, within the Markovian limit. In practice, the superconducting circuit elements are usually part of a larger circuit, constructed with many other linear circuit elements, which along with their environment is assumed to be mimicked by the baths. We study the response to variation of the coupling strength of resource generation for thee system prepared in zero-resource initial states. 
We focus, in particular, on entanglement and quantum coherence as resources.
We quantify the entanglement generation power of coupled transmon qutrits, taking into account the maximum entanglement the system can generate and the time-scale over which the system can sustain a significant entanglement. We identify the optimal initial separable states leading to maximum entanglement generating power. 
\end{abstract}

\maketitle

\section{Introduction}
In recent years, quantum information processing has become an important field of research due to its wide applications in quantum techniques such as computational speedup~\cite{Deutsch1992,Shor1997,Grover1997} and cryptographic security~\cite{Gisin2002}. Quantum computers can  potentially be realized with trapped ions~\cite{Leibfried2003,Blatt2012}, superconducting qubits~\cite{You2005,You2011,Nation2012,Wendin2017,Gu2017,Krantz2019,Kockum2019,Kjaergaard2020}, photons~\cite{Wang2016,Huang2017,Wang2018,Huang2018,Wang2019}, silicon chips~\cite{Kane1998,He2019}, etc. 
The superconducting Josephson junction qubit~\cite{Makhlin2001,Devoret2003,Devoret2004} is a leading candidate for the experimental realization of quantum computers. 
These superconducting systems derived from Josephson junctions, separated or joined by carefully chosen circuit elements, can primarily be divided into three subclasses according to their degree of freedom exploited to  realise the qubit, viz. charge~\cite{Bouchiat1998,Nakamura1999}, flux~\cite{Friedman2000,van_der_Wal2000}, and phase~\cite{Martinis2002}. 
But short coherence times pose a major challenge in extracting the optimal performances from such systems.

Transmon, a superconducting qubit,  has been proposed as a promising physical substrate for overcoming this limitation~\cite{Koch2007}. The structure of a transmon qubit is very similar to the Cooper pair box (CPB) qubit~\cite{Bouchiat1998} shunted with a large capacitor, and its ratio of Josephson energy to charging energy, $E_J/E_C$, lies in between that of charge and phase qubits. Two important parameters of a transmon qubit, viz., anharmonicity and charge dispersion of the energy levels, are usually determined by this ratio. The anharmonicity should be sufficiently large in order to prevent qubit operations from exciting other transitions in the system. Consequently, the charge dispersion needs to be reduced to minimize  the system's sensitivity to charge noise due to change in gate charge and stray electric fields~\cite{Ithier2005}. In a transmon qubit, the charge dispersion of the energy levels decreases exponentially with $E_J/E_C$ and the anharmonicity of the same reduces algebraically with a slow power law of $E_J/E_C$~\cite{Cottet2002}. Hence, in order to obtain the suitable operating regime of a transmon qubit, the ratio $E_J/E_C$ has to be chosen properly. For experimental realization of transmon qubits, see~\cite{Houck2009,Novikov2013,Peterer2015,Hu2017,Luthi2018,Patterson2019,Place2021,Wang2021,Li2021,Antony2021,Krause2022,Hertel2022,Majumder2022,Zemlicka2206}.

Entanglement \cite{Horodecki2009,Guhne2009,Das2019} and quantum coherence \cite{Aberg2006,Baumgratz2014,Winter2016,Streltsov2017} are two important resources for quantum information processing. The presence of environment influences the effective operating time of a transmon, resulting in a shorter coherence time. 
The coherence times of transmon qubits have been rigorously studied in previous literature in presence of various Markovian~\cite{Yang2021} and non-Markovian~\cite{Tuorila2019,Babu2021} environments. Entanglement between coupled transmon qubits has also been studied~\cite{DArrigo2012,Ohm2017,Deng2017,Dickel2018,Ye2018,Katabarwa2018,Hurant2020,Salmanogli2021,Cervera2022,Maiani2022}.

In this work, we consider two capacitively-coupled transmon qutrits, locally surrounded by harmonic oscillator baths. This setup provides not only a realistic setup, but also, interestingly, can self-generate quantum resources in the system, evolving from zero-resource initial states.
We also show enhancement in the generation of the quantum resources with increasing coupling strength. 
The maximum attainable as well as the long-time values of the quantum resources generated by the system increases almost linearly with the increase in the strength of capacitive coupling, restricting ourselves to the regime of weak to moderate coupling strength between the subsystems . 
We study the entanglement-generating power of such a coupled transmon setup, which depends on the maximum entanglement the system can generate, and the time-scale over which the system can sustain this self-generated entanglement to a significant amount. Furthermore, we find the optimal initial separable states of the composite system providing the maximum entanglement generating power, for small strengths of capacitive coupling between the two transmon qutrits.

The rest of the paper is arranged as follows. In Sec.~\ref{sec:2}, we describe the system under study. In Sec.~\ref{sec:3}, we discuss measures of the quantum resources that we have studied in this work. 
Section~\ref{sec:4} presents the results of quantum resource generation in the coupled-transmon system, for certain classes of paradigmatic initial states. 
In Sec.~\ref{sec:5}, we define the maximum entanglement-generating power of the system, and compute the optimal initial separable states that can generate maximum entanglement in presence of Markovian baths for fixed sets of system and bath parameters. A conclusion is presented in Sec.~\ref{sec:6}.

\section{Transmons coupled through charge-charge interaction }
\label{sec:2}

Structurally, a transmon is close to a  Cooper-pair box~\cite{Devoret2003}, and it can be constructed by two superconducting islands (or one island and ground), coupled through a Josephson junction and isolated from all other elements of the circuit~\cite{Koch2007}.
The system is described by a pair of canonically conjugate quantum operators constructed from the number of Cooper pairs transferred through the junction and the phase across the same.
The charge-based systems are sensitive to stray electric field noise. This unwanted situation can be largely overcome by putting the Cooper pair box (CPB) in the ‘transmon’ regime, where the Josephson tunneling energy dominates over the Coulomb charging energy \cite{Koch2007, ctn_jj4,ctn_jj5,Wendin2017, Martinis2002, ctn_jj_16, ctn_jj_18}.

The effective Hamiltonian of a CPB in the transmon regime can be written as~\cite{Devoret1997,Koch2007}
\begin{equation}
 H_{tr}=4E_C\hat{n}^2 - E_J\cos\hat{\phi},
\end{equation}
where $E_J=I_0\Phi_0/ 2\pi$ is the Josephson energy and $E_C=\frac{e^2}{2C}$ is the charging energy. Here we ignore the offset (or ‘gate’) charge given that the system is not sensitive to the offset charge in the transmon regime. \( \hat{n}=\hat{Q}/2e \) is the reduced charge operator ($2e$ being the charge content of a Cooper pair) and $\hat{\phi}={2\pi \Phi}/{\Phi_0}$ is the phase ($\Phi_0={h}/{2e}$ being the flux quanta) operator for the number of Cooper pairs and the phase across the Josephson junction respectively. Here, $h$ is the well known Planck's constant. $I_0$ is the maximum current allowed through the Josephson junction maintaining the superconducting state and $C$ is the total capacitance of the transmon circuit to its environment . The current ($I$) and flux ($\Phi$) across the Josephson junction are connected by the following Josephson relation: 
\begin{equation}
    I=I_0\sin{(2 \pi\Phi/\Phi_0)}.
\end{equation}

Quantization of this system is approached by constructing the ladder operators in terms of the zero-point fluctuations of charge and phase as
\[ \hat{n}=in_{zpf}(\hat{c}^{\dagger}-\hat{c}) \quad \mbox{and} \quad
   \hat{\phi}=\phi_{zpf}(\hat{c}^{\dagger}+\hat{c}),\]
with $n_{zpf}=\left({{E_J}/{32E_C}}\right)^\frac{1}{4}$ and $\phi_{zpf}=\left({{2E_C}/{E_J}}\right)^\frac{1}{4}$.
The operator \( \hat{c}=\sum_{j} \sqrt{j+1}\ket{j}\bra{j+1} \) is the transmon annihilation operator and the $\hat{c}^{\dagger}$ is the corresponding transmon creation operator. Note that the action of $\hat{n}$ provides integer values and changes by ±1 when a Cooper pair tunnels through the Josephson junction. In the transmon regime, \(E_J/E_C\gg1\), and hence \(\hat{\phi}\ll1\), providing the opportunity to simplify the Hamiltonian by expanding $\cos{\hat{\phi}}$ in Taylor series and then approximating the expression by ignoring small higher order contributions. If we retain terms upto $2^{\text{nd}}$ order in $\hat{\phi}$ only, and then express the canonical operators in terms of the  raising/lowering operators, the transmon Hamiltonian is approximated as
\begin{equation}
    \label{Etr}
    H_T \approx \sqrt{8E_CE_J}\Big(\hat{c}^{\dagger} \hat{c} +\frac{1}{2}\Big) -\dfrac{E_C}{12}(\hat{c}^{\dagger}+ \hat{c})^4 - E_J.  
\end{equation}

This Hamiltonian resembles that of a quantum  oscillator in a harmonic potential modified by a comparatively smaller quartic potential. This quartic term is again treated perturbatively in the transmon regime ($E_J \gg  E_C$) in a similar fashion.

In this work, we consider a pair of coupled transmon qutrits. Thus, here we concentrate on three lowest energy states of a multilevel transmon. We study the global quantum coherence of the combined system and the bipartite entanglement between the qutrits. The two transmons are coupled via an electric field.
This type of coupling occurs when the individual transmon circuits are coupled through an electric field via a capacitor. This is particularly plausible when the impedance of the source circuit is high \cite{Koch2007, othr_9, ctn_jj_10}. This kind of interaction between two transmons can be modelled as~\cite{Krantz2019,ctn_jjcouple_20}
\begin{equation}
\label{intrn}
H_{int} = \hbar \gamma \hat{n}_1 \hat{n}_2=- \hbar
\frac{\gamma}{\sqrt{32}}  
\left({\dfrac{E_{J_1}E_{J_2}}{E_{C_1}E_{C_2}}}\right)^\frac{1}{4}
(\hat{c}_1^{\dagger}-\hat{c}_1)(\hat{c}_2^{\dagger}-\hat{c}_2).
\end{equation}

Here we neglect the charge offset terms, since the transmon regime provides us with the charge-insensitive (insensitive to gate charge) regime to work in. The coupling constant \( \gamma \) is taken to be roughly one order of magnitude smaller than the ground state energy,
in view of the fact that this kind of coupling is often weak compared to the energy-scales of the individual systems. 
The Hamiltonian of this composite two-transmon system is taken as 
\begin{equation}
    H_{s}=\sum_{i=1}^2 \Big[\hbar \omega_{0_i} \Big(\hat{c}_i^{\dagger} \hat{c}_i +\frac{1}{2}\Big) -\dfrac{E_{C_i}}{12}        (\hat{c}_i^{\dagger}+ \hat{c}_i)^4\Big]+ H_{int},
    \label{hamil_to}
\end{equation}

where \(\omega_{0_i}=\sqrt{8E_{C_i}E_{J_i}} / \hbar \) ($i=1,2$).
Here, we have dropped the constant contributions coming from $E_{J_1}$ and $E_{J_2}$ (see Eq.~(\ref{Etr})) as these terms just give a constant shift to all energy values. In all the further discussions of this paper, we have taken $E_{J_1}=E_{J_2} \equiv E_J$,  $E_{C_1}=E_{C_2}\equiv E_C$ and $\omega_{0_1}=\omega_{0_2}\equiv \omega_0=\sqrt{8E_CE_J}$.

\section{Measures of quantum resources}
\label{sec:3}

\begin{figure*} 
\begin{framed}
\begin{minipage}{0.48\textwidth}
 \centering 
     \includegraphics[width=\textwidth]{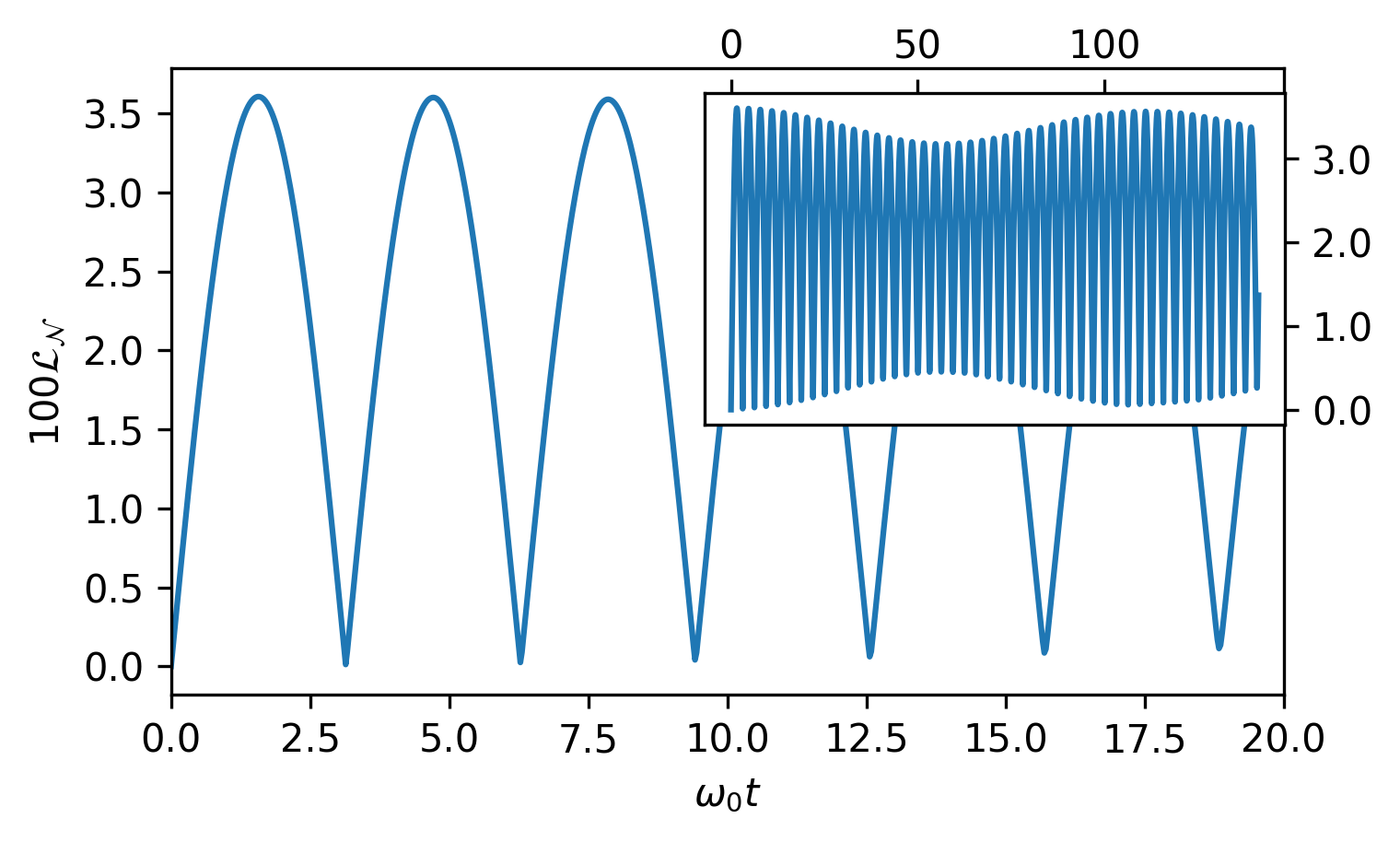}
    \centering
    \textbf{(a)}
\end{minipage}
\hfill
\begin{minipage}{0.48\textwidth}
 \includegraphics[width=\textwidth]{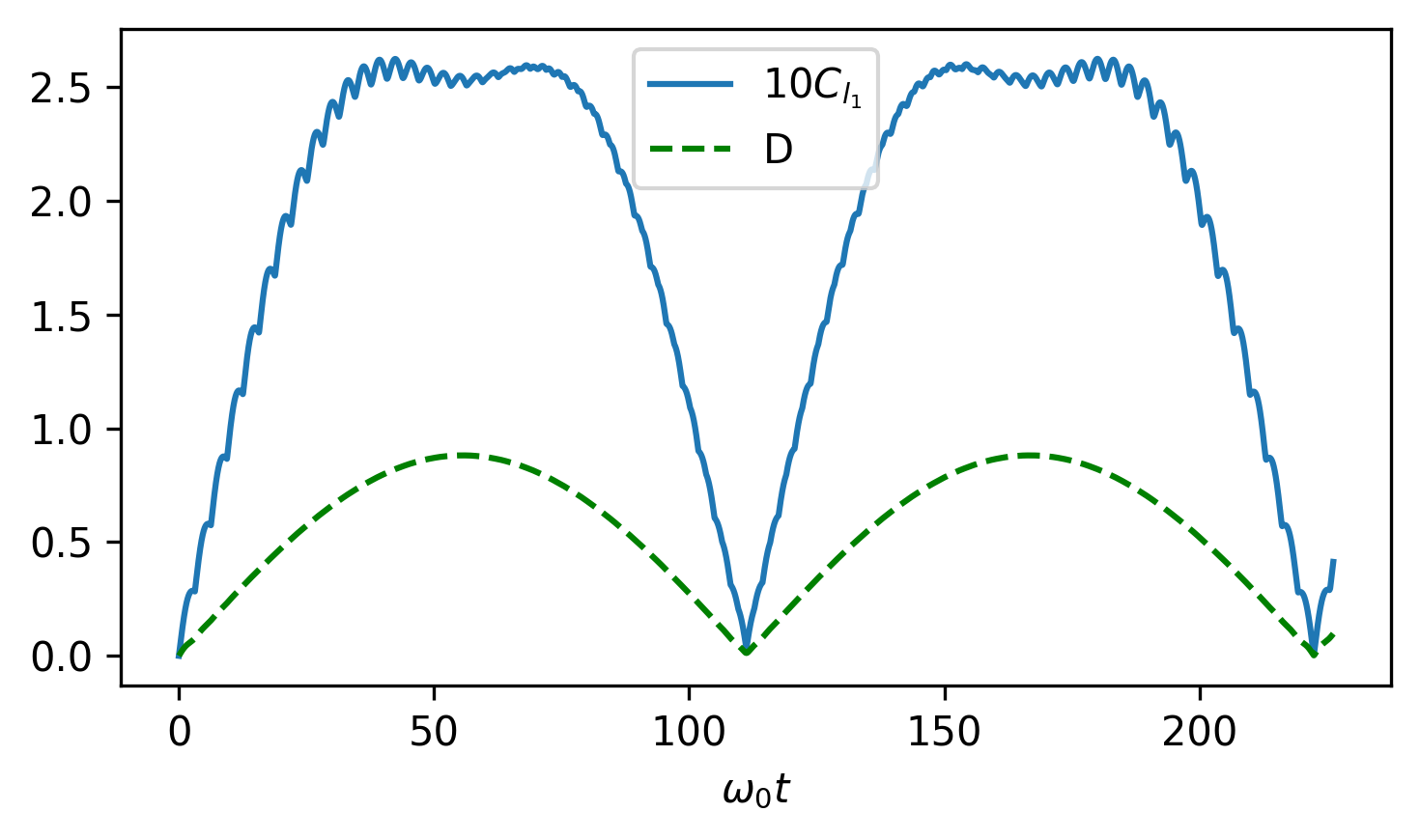}
    \centering
    \textbf{(b)}
\end{minipage}
 \caption{Time dynamics of the quantum resources for a pair of coupled transmon qutrits isolated from the surroundings. Here we plot the variation of the logarithmic negativity of entanglement with time for the composite system of interacting transmon qutrits described in Eq.~\eqref{hamil_to} in panel (a). The initial state of the system is \( \ket{00}\). The long time behavior of entanglement for this case is presented in the inset. The $l_1$-norm of quantum coherence of the system with the initial state \( \ket{11}\) is depicted in panel (b) with blue line. The green dashed line in panel (b) presents the trace distance between the time-evolved state and the initial state \( \ket{11}\). The ratio of the Josephson energy to the charging energy for both the subsystems is taken to be \(E_J/E_C=100\). The coupling-strength between the subsystems is chosen as \(\gamma=0.2 E_C/\hbar\). The quantities $\mathcal{L}_{\mathcal{N}}$ and $C_{l_1}$ demonstrated along the vertical axes of both the panels are in ebits and bits respectively and the quantity $D$ depicted by green dashed line in panel (b) is dimensionless. The quantities plotted along the horizontal axes are dimensionless as well.} 

\label{fig:1}
\end{framed}
\end{figure*}

Our aim in this paper is to investigate the dynamics of various quantum resources, e.g. quantum coherence and quantum entanglement, in the system studied. 
Of these two resources, quantum coherence is a basis-dependent quantity, while entanglement is basis-independent. In this work, the $l_1$-norm of quantum coherence and logarithmic negativity are chosen as the quantum coherence and entanglement measures respectively. \\
\\
\textbf{$\mathbf{l_1}$-norm of quantum coherence:} The $l_1$-norm of quantum coherence of an arbitrary \(d\)-dimensional state $\rho$, possibly mixed, is defined as the sum of the moduli of the off-diagonal terms, when the state is expressed in a fixed reference basis \cite{Aberg2006,Baumgratz2014,Winter2016,Streltsov2017}. Let us consider a $d$-dimensional Hilbert space $\mathbb{C}^d$ and fix a basis of it, $\{\ket{i}\}$, for $i=1,2,\ldots,d$, as the reference basis. Then, the $l_1$-norm of quantum coherence of an arbitrary \(d\)-dimensional quantum state \( \rho=\sum_{i,j} p_{ij}\ket{i}\bra{j}\), with respect to the basis $\{\ket{i}\}$, is given by  \(
\tilde{C}_{l_{1}}(\rho)= \sum\limits_{i\neq j}|p_{ij}|
\). 
The maximum value of the $l_1$-norm of quantum coherence of a state on \(\mathbb{C}^d\) is given by \(\tilde{C}_{l_{1_{\text{max}}}} ={(d^2-d)}/{d} = d-1\). So, in order to normalise the obtained value of quantum coherence, \(
\tilde{C}_{l_{1}}(\rho)\) is re-scaled by \(\tilde{C}_{l_{1_{\text{max}}}}\). 
In this work, we consider the states on the Hilbert space constructed from the the three lowest-lying states of each of the two coupled transmon systems. The two-transmon states are therefore defined on $\mathbb{C}^3\otimes \mathbb{C}^3$, and so, \(\tilde{C}_{l_{1_{\text{max}}}}=8\). Hence, the expression of the $l_1$-norm of quantum coherence, taken in all further considerations of this paper, is given by 

\begin{equation}
C_{l_{1}}(\rho)=\dfrac{1}{8} \sum\limits_{i\neq j}|p_{ij}|.
\end{equation}

The maximum value of $C_{l_{1}}(\rho)$, therefore, reaches to unity in the $\mathbb{C}^3\otimes \mathbb{C}^3$ system.

\textbf{Logarithmic negativity of entanglement:} One of the most popular measure of bipartite entanglement is the logarithmic negativity~\cite{Vidal2002,ctn_ent7}. Suppose, $\rho$ is a two-party ($A$ and $B$) density matrix and $\rho^{T_A}$ is the partial transpose of $\rho$ on the subsystem $A$. The negativity of the density matrix $\rho$ is defined as
\begin{equation}
    \mathcal{N}(\rho)=\frac{||\rho^{T_A}||_1-1}{2}.
\end{equation}
Here, the trace norm of an operator $A$ stands for $||A||_1=\text{Tr}\big(\sqrt{A^{\dagger}A}\big)$. This $\mathcal{N}(\rho)$ presents the absolute value of the sum of negative eigenvalues of $\rho^{T_A}$~\cite{Zyczkowski1998,Zyczkowski1999}. The logarithmic negativity of \(\rho\) is now defined as 
\begin{equation}
    \mathcal{L_N}(\rho) = \log_{2}(2 \mathcal{N}(\rho)+1).
\end{equation}

\section{Resource generation in a pair of  coupled transmon qutrits for paradigmatic initial states}
\label{sec:4}

In this section we investigate the variation of quantum coherence and entanglement for different initial states of a pair of interacting transmon qutrits described by the Hamiltonian $H_{s}$ (see Eq.~(\ref{hamil_to})). We consider two different situations.
First, for completeness, we consider the unitary evolution of the isolated system solely governed by the Hamiltonian \(H_s\).
Next we investigate the scenario where each transmon qutrit is locally connected to a bosonic bath within Markovian limit. We consider weakly coupled harmonic baths, where the coupling between a transmon and a bath is weak and the baths are infinitely large, having a continuously distributed energy spectrum, for the validation of the Born-Markov approximations~\cite{Petruccione,Alicki,Rivas,Lidar}.

\subsection{Coupled transmons isolated from the environment}
\label{sec3A}

We now present the time dynamics of quantum resources in the system described, isolated from any environmental effects.
 In this work, we have fixed the \(E_J/E_C\) ratio to 100, in order to confine the subsystems in the transmon regime. We have considered the value of \(\hslash\gamma\), the coupling strength between the subsystems, to be one order of magnitude smaller than the capacitance energy to remain in the weak coupling limit. We took this value to be 0.2 for further studies. The anharmonicity in the energy spectrum is quantified by difference in the consecutive excitation energies of this system, given by \(E_{12}-E_{01} \), where energy difference between two eigenstates is denoted by \(E_{mn}=E_n-E_m\). The aforementioned parameter space considered in this work leads to the value of ground state oscillator energy to be \(20\sqrt{2} \hslash\), and the anharmonicity in the first three energy levels to be \((-E_c)\), keeping terms up-to \(1^{\text{st}}\) order energy-correction.
We take the basis $\mathbb{B}=\{\ket{ij}\}$, with $i$ and $j$ running from $0$ to $2$, as the reference basis for evaluation of the $l_1$-norm of quantum coherence in this paper.

So, in all the succeeding discussions, by ``$l_1$-norm of quantum coherence", we will actually mean the global $l_1$-norm of quantum coherence of the composite two transmon system with respect to the reference basis $\mathbb{B}$.

From Fig.~\ref{fig:1}-(a), one observes a periodic collapse and revival of entanglement with time when the system is initiated in the pure product state $\ket{0 0 }$. We also study the time dynamics of global $l_1$-norm of coherence, along with the trace distance~\cite{Ma2008} and fidelity~\cite{Ma2008,Jozsa1994,Dodd2002}, of the time evolved state from the initial one. The trace distance $D$ and fidelity $F$ between two arbitrary states $\rho$ and $\sigma$ are defined as, respectively,

\begin{eqnarray}
    &&D(\rho, \sigma) = \frac{1}{2} ||\rho-\sigma||_1, \quad \text{and}\\
    &&F( \rho, \sigma )= \left( \text{Tr} \sqrt{ \sqrt{\rho} \sigma \sqrt{\rho} } \right)^2.
\end{eqnarray}
We observe that these quantities also vary in a qualitatively similar fashion like the entanglement, but with different amplitudes, and in some cases with a phase-shift of \(\pm \pi\).

In Fig.~\ref{fig:1}-(b), we demonstrate the time variation of the \(l_1\)-norm of coherence and trace distance under the same unitary evolution
when the system is initially prepared in the $\ket{11}$ state. As in the previous case (panel (a)), and as it may be expected, oscillatory behavior in time persists. The entanglement generated from the composite initial state $\ket{11}$ changes in a similar manner as the coherence of the same initial state (see panel (b) blue line), the only difference being the absence of the small ``eddy" oscillations present in the coherence dynamics.
Various other initial separable states, e.g. $\ket{++}$, have been considered. The qualitative features remain similar irrespective of the choice of initial state from the separable set.
The amplitude profile of the oscillatory dynamics of  entanglement and quantum coherence changes in a periodic manner, as can be seen from the inset of
Fig.~\ref{fig:1}-(a). Noticeably, the frequency and maximum amplitude of this envelope increases rapidly with increasing coupling strength.
In short, the measures of the quantum resources studied in this paper, behaves in a qualitatively similar fashion under unitary evolution, when the system starts from  separable states.


On the contrary, entanglement remains almost constant in time  
when the system starts from the entangled state , \((\ket{01}+\ket{10})/\sqrt{2}\).
We find that  this result remains the same for some other  entangled states , e.g., \((\ket{12}+\ket{21})/\sqrt{2}\), as well. 
However, the global quantum coherence of the system still maintains the oscillatory behaviour, which is very similar to the entanglement in Fig. \ref{fig:1}-(a) for these entangled states.

Therefore, from the observations of the unitary evolution of a pair of coupled transmons, we can infer that one can generate sufficient quantum resources by a coupled transmon system, when the composite system is isolated from its surroundings. These scenarios are far from the actual situations, as in reality, the described transmon qutrits are never isolated, but are connected to other circuit elements as well. These circuit elements can be considered as environments attached with the individual qutrits. Environments may cause substantial adverse effects on the time dynamics of the quantum resources for the system under consideration. In the succeeding sections, we will study the effects of environments on this ``ideal" coupled-transmons setup.

\subsection{Dynamical equation in presence of Markovian baths}

\begin{figure*}
\begin{framed}
    
\begin{minipage}{0.47\textwidth}
\includegraphics[width=\textwidth, height=5cm]{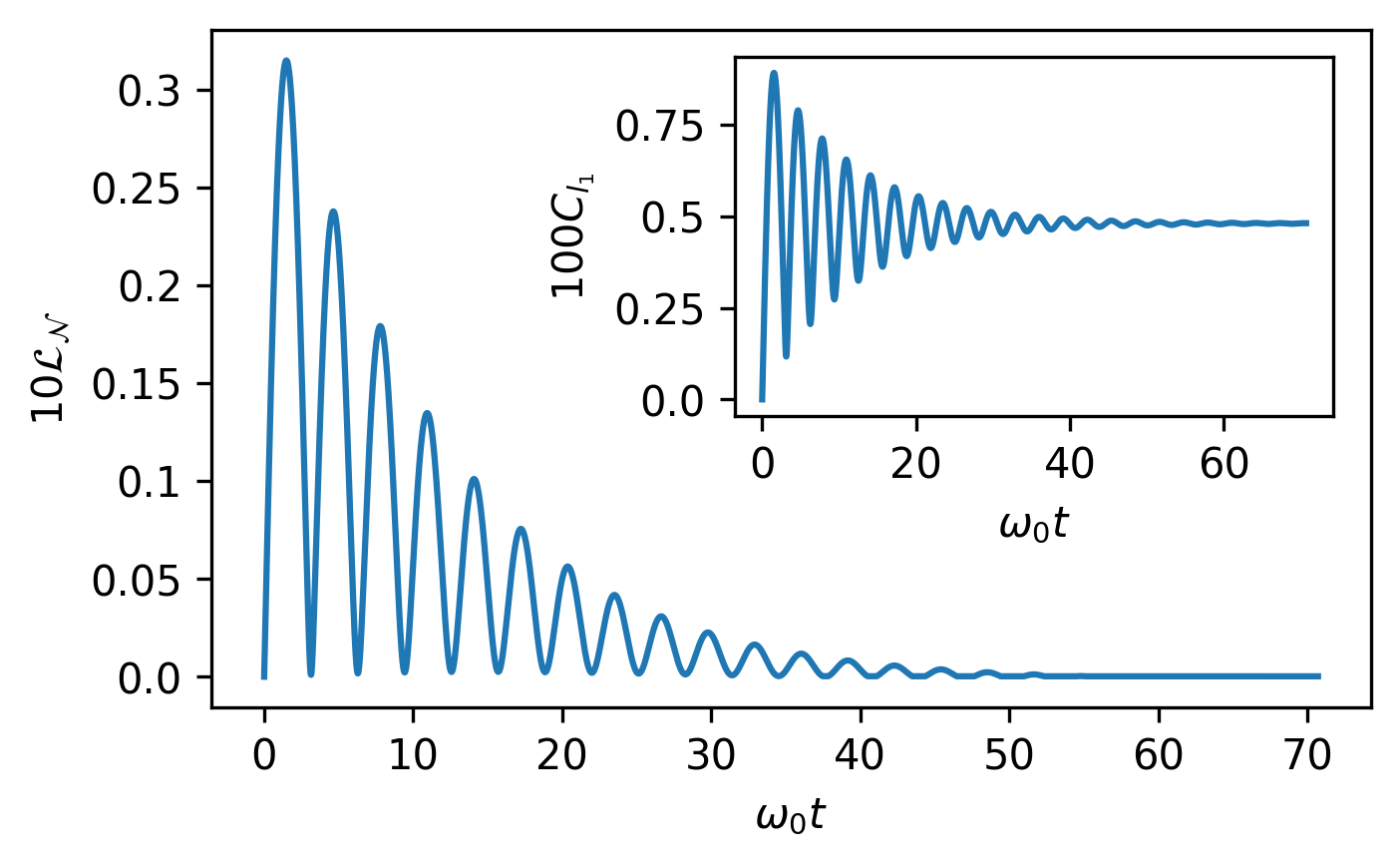}
\centering
\textbf{(a)}
\end{minipage}
\hfill
\begin{minipage}{0.47\textwidth}
\includegraphics[width=\textwidth, height=5cm]{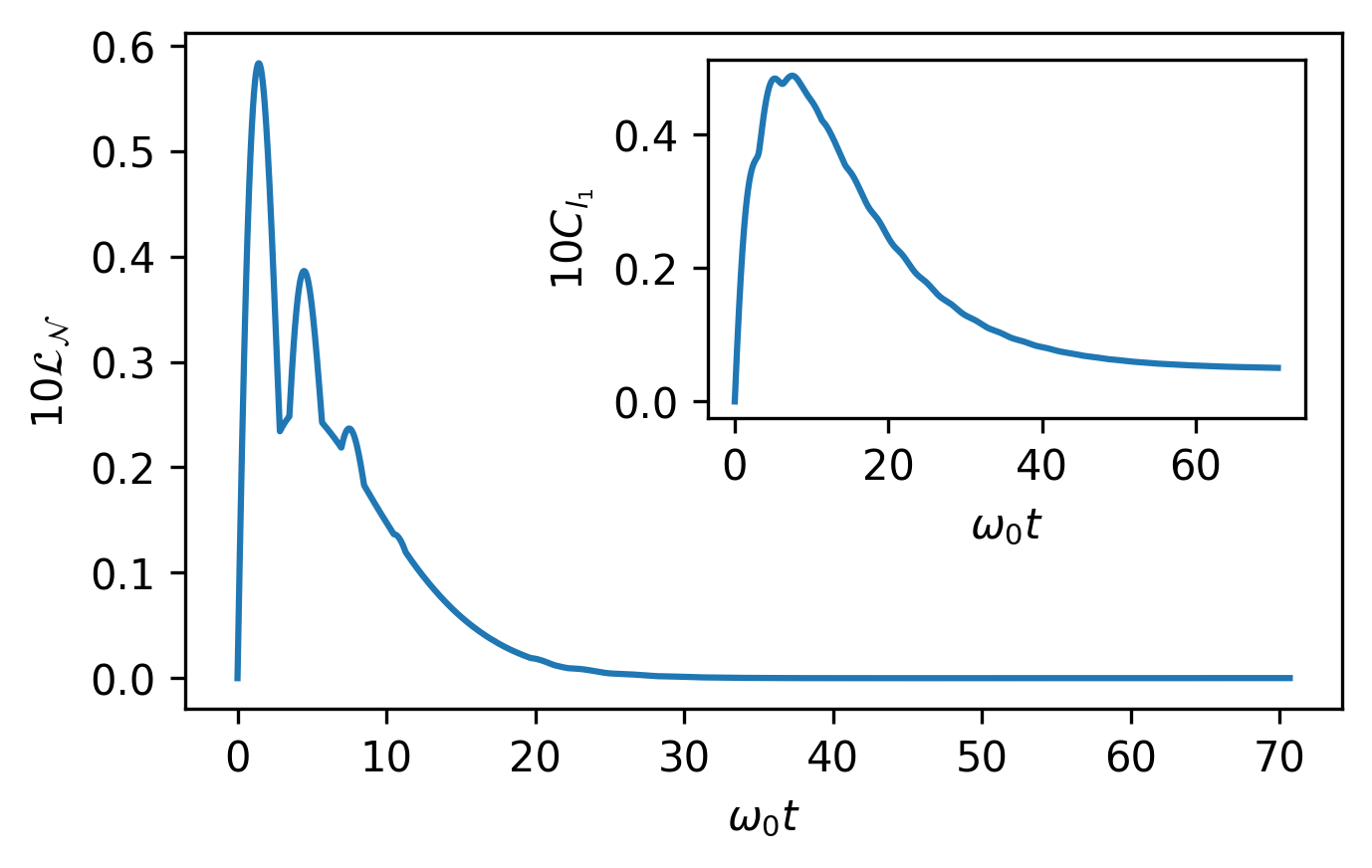}
 \textbf{(b)}
 \end{minipage}
\caption{Time dynamics of the quantum entanglement and quantum coherence for initial separable states of the coupled transmon qutrits locally immersed in Markovian bosonic baths. Here we demonstrate the logarithmic negativity of entanglement between the qutrits of the system in the main figures and the global $l_1$-norm of coherence of the system in the insets for the initial states (a) $\rho_s(0)= \ket{00} \bra{00}$ and  (b) $\rho_s(0)= \ket{11} \bra{11}$. 
The system parameters are same as in Fig.~\ref{fig:1}. We have taken \(\kappa_1=\kappa_2=\omega_{01}/20\), and the temperature of the baths are chosen to be \(\beta_1=\beta_2=5/(\hbar \omega_{01})\). The quantities $\mathcal{L}_{\mathcal{N}}$ and $C_{l_1}$ plotted along the vertical axes are in ebits and bits respectively. The quantities presented along the horizontal axes are dimensionless.} 
\label{rr_bb_u} 
\end{framed}

\end{figure*}

Quantum systems are generally susceptible to environmental effects, and hence it is important to study the influence of environments on the generation of quantum resources in coupled transmon qutrits. In that spirit, here we consider the system described by the Hamiltonian $H_s$, in presence of two independent harmonic oscillator baths, separately coupled to the two transmon qutrits locally. As mentioned before, the system-bath coupling is taken to be weak for the validation of Born-Markov approximations~\cite{Petruccione,Alicki,Rivas,Lidar}.
\begin{equation}
    H_{B_i}=\sum_j \hbar \nu_j^i b_j^{i\dagger} b_j^i,
\end{equation}
for $i=1$ and $2$. $\nu_j^i$ is the frequency of the $j^{\text{th}}$ mode of the $i^{\text{th}}$ bath. $b_j^i$ ($b_j^{i \dagger}$) is the bosonic annihilation (creation) operator of $i^{\text{th}}$ bath corresponding to $j^{\text{th}}$ mode. The oscillator modes interacts with the system through the net content of cooper pairs across the junction, i.e. the reduced charge (\(\hat{n}\)) parameter of the system. This introduces system-bath interaction of the following form~\cite{Tuorila2019}
\begin{equation}
    H_{SB} = \sum_{i=1}^2\hbar q^i \sum_{j} g_{j}^i (\hat{b}_j^{i\dagger} + \hat{b}_j^i),
\end{equation}
with $g_{j}^i$ being the coupling constant having the unit of frequency, for tuning the interaction strength between $i^{\text{th}}$ system and $j^{\text{th}}$ mode of the $i^{\text{th}}$ bath. Here \(q^{1} = \big( \sum_{l,m} \ket{l}\bra{l} \hat{n}_{1}\otimes \mathbb{I}_3\ket{m}\bra{m} \)\big) and \(q^{2} = \big( \sum_{l,m} \ket{l}\bra{l} \mathbb{I}_3 \otimes \hat{n}_{2} \ket{m}\bra{m} \)\big), i.e., the reduced charge (number of cooper pairs) operators expressed in the eigen-basis  of system-Hamiltonian $H_s$, $\mathbb{I}_3$ being the identity matrix on the three dimensional Hilbert space.
Hence, the total Hamiltonian of the composite system-bath setup, with harmonic oscillator baths coupled to each transmon qutrits locally in the circuit, reads 
\begin{equation}
H_{total}=H_s+\sum_{i=1}^2 H_{B_i}+H_{SB}.
\label{eq_h_tot}
\end{equation}

In presence of these Markovian bosonic baths the system undergoes a open system dynamics governed by the Gorini-Kossakowski-Sudarshan-Lindblad (GKSL) master equation given by~\cite{Petruccione,Alicki,Rivas,Lidar}
\begin{equation}
    \frac{d\rho_{s}(t)}{dt}=-\frac{i}{\hbar}\Big[H_s,\rho_s(t)\Big]+\sum_{i=1}^{2}\mathcal{D}_i(\rho_s(t)).
\end{equation}
Here $\rho_s(t)$ is the composite two transmon state at time $t$ after tracing out the Markovian baths. $\mathcal{D}_i(\rho_s(t))$ is the dissipative term coming from the interaction between the $i^{\text{th}}$ system and $i^{\text{th}}$ bath, presented as
\begin{eqnarray}
&&D_i(\rho_s(t))=\frac{1}{2} \sum_{\omega_{nm}>0} S_i(\omega_{nm})\Big[2\Pi^i_{nm}\rho_s(t)\Pi^{i\dagger}_{nm}\nonumber\\
&&\phantom{ami}-\big\{\Pi^{i\dagger}_{nm}\Pi^i_{nm},\rho_s(t)\big\}\Big] +\frac{1}{2} \sum_{\omega_{nm}>0} S_i(-\omega_{nm})\nonumber\\
&&\phantom{kbe a}\times \Big[ 2\Pi_{nm}^{i\dagger}\rho_s(t)\Pi^i_{nm}-\big\{\Pi^i_{nm}\Pi^{i\dagger}_{nm},\rho_s(t)\big\}\Big]\nonumber\\
&&+\frac{1}{2}\sum_n S_i(0)\Big[2\Pi^i_{nn}\rho_s(t)\Pi^i_{nn}-\big\{\Pi^i_{nn}\Pi^i_{nn},\rho_s(t)\big\}\Big],\nonumber\\
\end{eqnarray}
where $\omega_{nm}=\omega_m-\omega_n$. Here $\omega_n$ denotes the eigen-frequency of the system and $\Pi^i_{nm}$ are the Lindblad operators, for $i=1$ and $2$, where $\Pi^i_{nm}=q^i_{nm}\ketbra{n}{m}$ and $q^i_{nm}=\bra{n}q^i\ket{m}$. The transition rate, $S_i(\omega)$, turns out to be 
\begin{equation}
S_i(\omega) = \dfrac{J_i(\omega)}{1-e^{-\hbar\beta_i\omega}},
\end{equation}
where $\beta_i={1}/{k_BT_i}$. Here $T_i$ denotes the temperature of the $i^{\text{th}}$ bath and $k_B$ is the Boltzmann constant.
$J_i(\omega)$ represents the spectral density function of the harmonic oscillator baths, which we have chosen to be Ohmic spectral density function. The functional form is given by
\begin{equation}
J_i(\omega)=\dfrac{\kappa_i\omega/\omega_{01}}{[1+(\omega/\Omega_{c_i})^2]^2},
\end{equation}
where \(\omega_{01}\) denotes the energy-difference between the first two levels of the composite transmon system described by the Hamiltonian $H_s$. We fix the drude cut-off \(\Omega_{c_1}=\Omega_{c_2}=50\omega_{01}\) in order to keep this cut-off frequency sufficiently large compared to the low-lying energy-levels of the system.
As a result, the system-bath interaction belongs to the weak-coupling regime $S(\omega_{nm})\ll |\omega_{nm}|$ ~\cite{Petruccione,Alicki,Rivas,Lidar}.
The system-bath coupling constants are considered to be \( \kappa_1=\kappa_2=\omega_{01}/20 \) for subsequent computations throughout this work, so that the Markovian approximations remain valid.


\subsection{Resource generation in presence of bosonic baths}
In Fig.~\ref{rr_bb_u}, we describe the time variations of logarithmic negativity of entanglement and $l_1$-norm of quantum coherence of the two transmon system considered in Eq.~(\ref{hamil_to}), initially starting from separable states. We observe that 
the entanglement between the two qutrits of the system, as well as the global quantum coherence of the system, initially increases quickly to a maximum value and then decays to a steady value in an oscillatory manner with decreasing amplitudes of oscillations. 
These intermediate oscillations between the maximum and the saturated value are more prominent when the system is initialized in the $\ket{00}$ state than the case when the system starts from an excited state $\ket{11}$ (panels (a)-(b) of Fig.~\ref{rr_bb_u}). 

\begin{figure}
\begin{framed}
    
\includegraphics[width=\columnwidth]{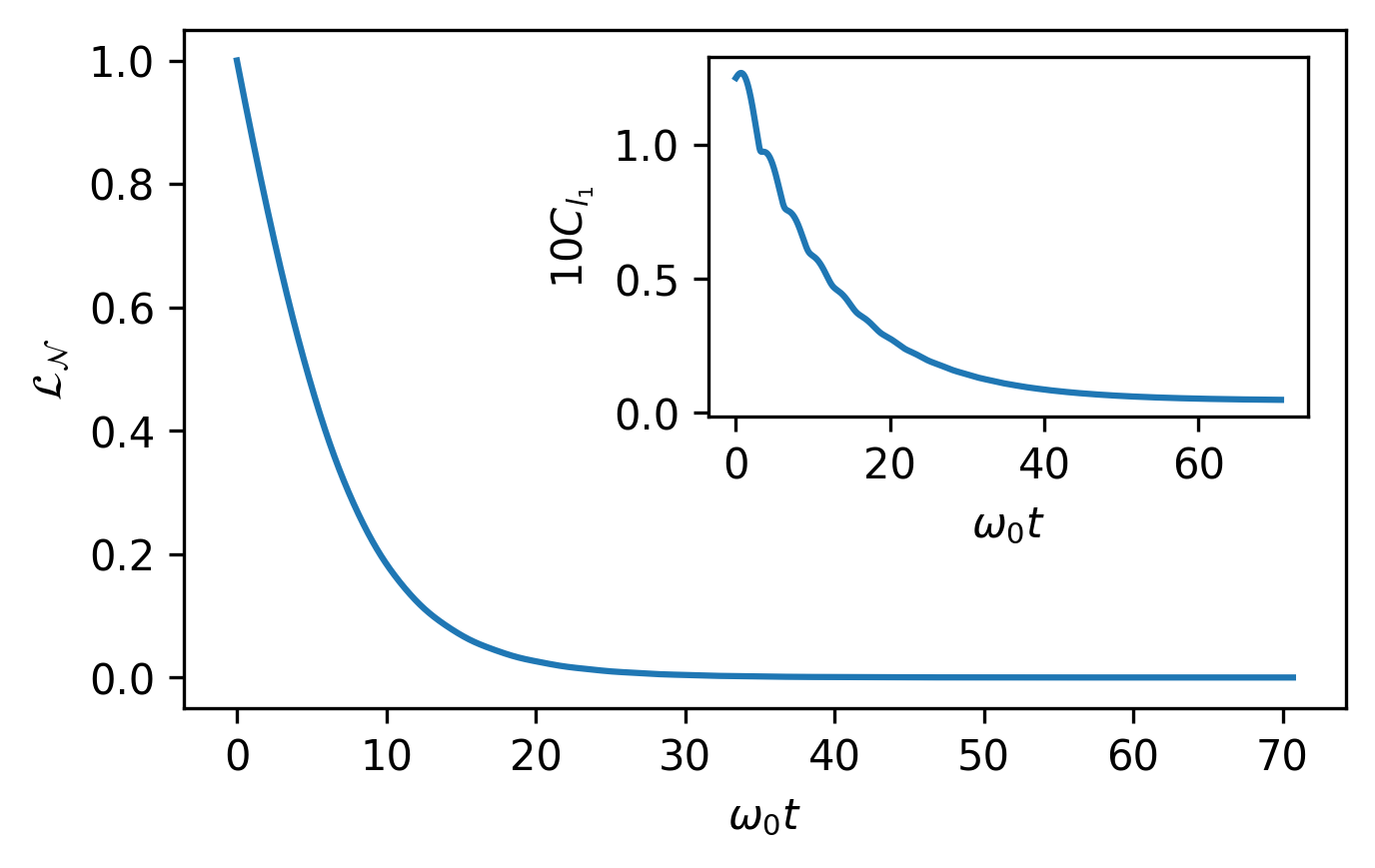} 
    \caption{Time dynamics of quantum entanglement and quantum coherence for an initial entangled state of the coupled transmon qutrits locally immersed in Markovian bosonic baths. Here we depict the time evolution of logarithmic negativity between the two qutrits (in the main figure) and the global $l_1$-norm of quantum coherence of the system (in the inset) in presence of Markovian baths for the initial state $\rho_s(0)=\ket{\psi_s(0)}\bra{\psi_s(0)}$, with $\ket{\psi_s(0)}=  (\ket{01}+\ket{10})/\sqrt{2}$. All other considerations are same as in Figs.~\ref{fig:1} and~\ref{rr_bb_u}.
    The quantities $\mathcal{L}_{\mathcal{N}}$ and $C_{l_1}$ plotted along the vertical axes are in ebits and bits respectively. The quantities presented along the horizontal axes are dimensionless.}
\label{ent_bath_1}
\end{framed}

\end{figure}

This behaviour is, however, distinctly different when the system is initially prepared in an entangled state. When we start from the entangled state $(\ket{01}+\ket{10})/\sqrt{2}$ for the each subsystem,  the entanglement and quantum coherence decays exponentially. See Fig.~\ref{ent_bath_1}. We also check for the state $(\ket{12}+\ket{21})/\sqrt{2}$. The nature of the two resources are qualitatively same.

In absence of any interaction in the system Hamiltonian i.e., for $\gamma=0$ , system only goes through decoherence. As a result, an initial system density matrix with nonzero quantum coherence decays to a zero resource state very soon, and a separable state remains separable. On the contrary, the system studied in this work, has a non-local interaction term $H_{int}$ in the  system Hamiltonian, arising due to the interaction between the transmon components present in the system.
This non-local interaction term works in favour of continuously feeding quantum resource in the system, whereas quantum resources leak out into the environment simultaneously, due to the Markovian baths. These two mechanisms act in opposition.In small initial times the quantum resources exhibit oscillatory nature as in the unitary case (see Fig.~\ref{fig:1}) since the decoherence caused due to the baths are very small then, which gradually build up to a substantial value. In the long time limit, the competition between these two mechanisms lead the system towards a steady state. Consequently, the quantum resources considered in this work attain steady values s well.

\begin{figure}
\begin{framed}
    
\includegraphics[width=\columnwidth]{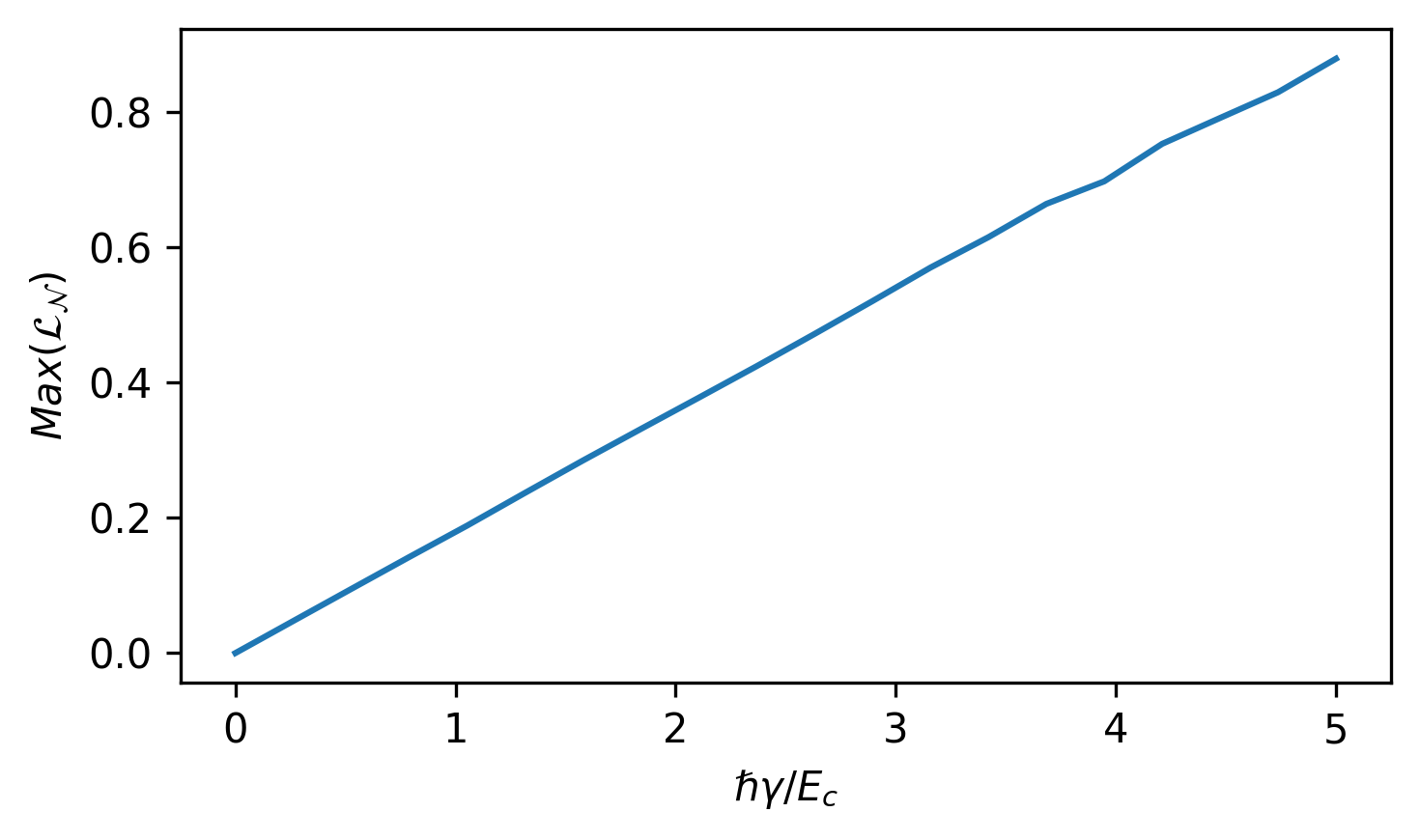}
    \caption{The variation of the maximum entanglement with varying capacitive coupling strength under the unitary evolution described by the Hamiltonian in Eq.~\eqref{hamil_to}. The initial state is chosen to be the separable state \( \rho_s(0)=\ket{00}\bra{00}\). All other considerations are same as in Fig.~\ref{fig:1}. The quantity plotted along the vertical axis is in ebits and the same plotted against the horizontal axis is dimensionless.}
    \label{fig:my_label}
    \end{framed}

\end{figure}

\subsection{Dependence of resource generation on capacitive coupling strength}

The capacitive coupling between the transmon qutrits can be implemented in real experimental setups conveniently. 
The experimental setup of the system studied in our work can be built with  two Josephson junctions of Josephson energy \(E_J\) , shunted with two capacitors \( C_1\) and \(C_2\), yielding a charging energy \(E_C\) in the individual circuits. When a capacitor with capacitance \(C_g\) is placed between the voltage nodes (with voltages \(V_1\) and \(V_2\)) of the two participating transmon circuits, we obtain the capacitive coupling. This yields a non-local interaction term in the Hamiltonian of the form \cite{Krantz2019}
$$ H_{I}= C_g V_1 V_2.$$
Circuit quantization in the limit of \(Cg \ll C_1,C_2\) yields an interaction term in the form of Eq.~\eqref{intrn}, where we can express the interaction parameter \(\gamma\) in terms of the circuit elements as 
\begin{equation}
    \hbar\gamma=4e^2 \dfrac{C_g}{C_1C_2}.
\label{g_int}
\end{equation}

The quantum resources exhibit an oscillatory nature and reach to a maximum value during the oscillation for the unitary evolution of the system described in Sec.~\ref{sec3A} (see Fig.~\ref{fig:1}). We observe that these maximum values of the bipartite entanglement and the global quantum coherence in such an isolated system of two capacitively coupled transmon qutrits increase with increasing strength of the interaction $\gamma$. The case for logarithmic negativity of entanglement is depicted in Fig.~\ref{fig:my_label} for increasing values of $\gamma$. The global $l_1$-norm of quantum coherence of the system also varies similarly with the transmon-transmon coupling strength $\gamma$.

\begin{figure*}
\begin{framed}

\begin{minipage}{0.64\columnwidth}
\hspace{-2.5em}    
\centering
\includegraphics[width=\textwidth]{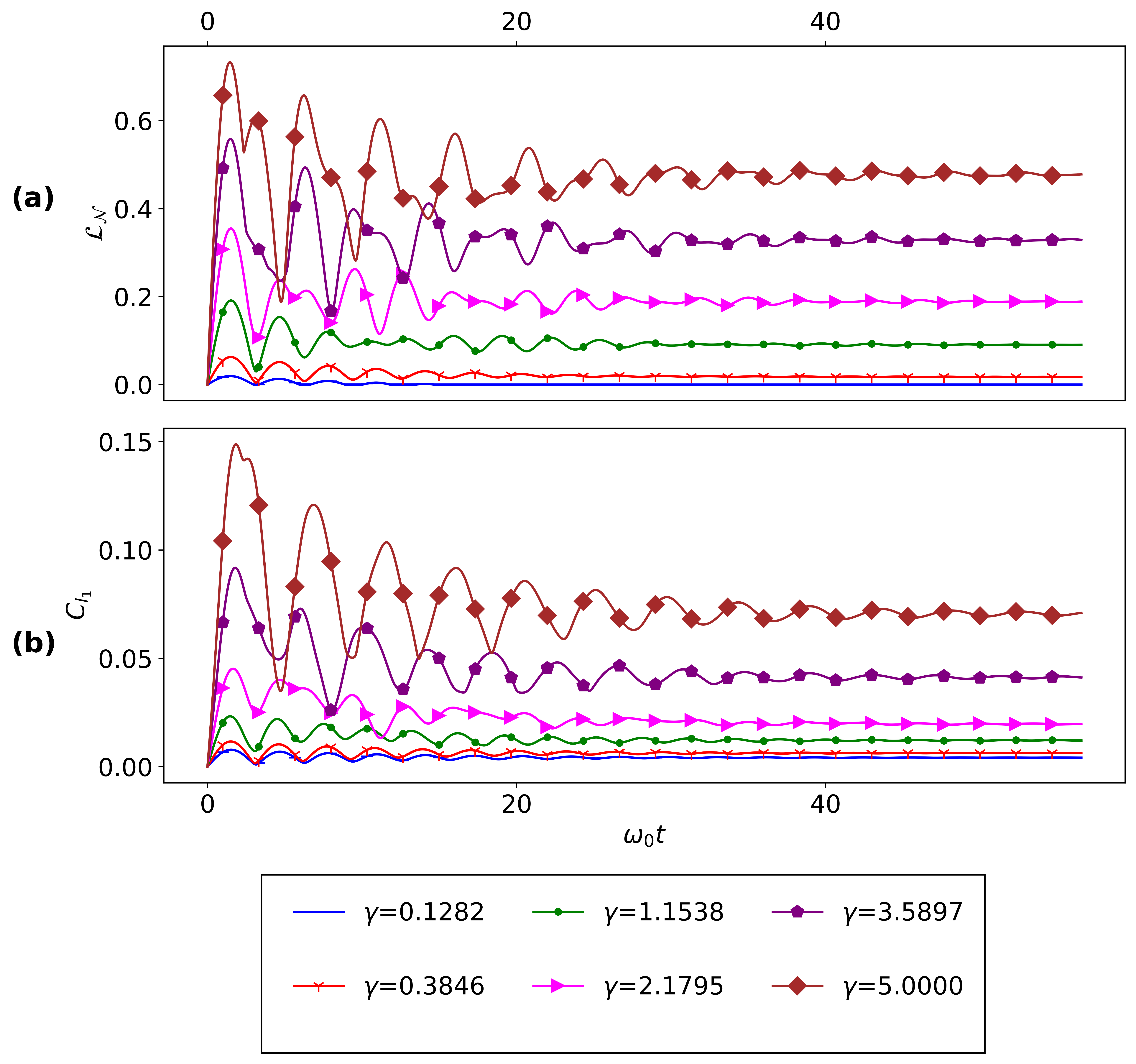} 
\end{minipage}
  \begin{minipage}{0.34\columnwidth}
 \caption{Quantum resource generation in the coupled transmons for different values interaction strength between the two capacitively coupled transmon qutrits, both immersed in bosonic baths as described by Eq.~\eqref{eq_h_tot}. Panel (a) describes the time variation of the logarithmic negativity of bipartite entanglement in the system, and panel (b) describes the time evolution of the $l_1$-norm of coherence of the system. The initial state is taken as  \(\ket{\psi_s(0)}=\ket{00}\), with zero quantum resource. The values of the interaction strengths \(\gamma\) are given in the legends in units of \(E_c/\hbar\). All other considerations are same as in Figs.~\ref{fig:1} and~\ref{rr_bb_u}. The quantities $\mathcal{L}_{\mathcal{N}}$ and $C_{l_1}$ plotted along the vertical axes are in ebits and bits respectively. The quantities presented along the horizontal axes are dimensionless.} 
\end{minipage} 
\label{cc_g} 
\end{framed}

\end{figure*}

We now scrutinize the dependence of resource generation on the coupling strength $\gamma$ for the coupled transmons setup immersed in Markovian bosonic baths, described in the preceding section. In Figs.~\ref{cc_g}((a)-(b)) we depict the time dynamics of logarithmic negativity of entanglement and the $l_1$-norm of quantum coherence, respectively,s
for different strengths of the capacitive coupling between the two transmon qutrits. 
Here we vary the interaction strength in a small regime so that the circuit quantization yielding the interaction term in the form mentioned in Eq.~\eqref{g_int} remains valid.
As can be seen from Fig.~\ref{cc_g}, that 
the maximum values of bipartite entanglement and quantum coherence, along with the long time steady state values of the same, increase with stronger coupling between the subsystems, and finally sustain a non-zero finite value. 
This demonstrates that, for initial states with zero quantum resources and even for very small nonzero interactions  in the form of Eq.~\eqref{intrn}, not only the maximum quantum resources in the system studied increases, but the decoherence time increases profoundly. This result shows that if we can create such coupled transmon circuits with tunable charge-charge interaction, that promises us control over the decoherence time of the system, which translates to improved practical implementations of the described system in future. 
Note that, we cannot increase the self-generated quantum resources indefinitely by choosing a higher value of $\gamma$. In the range from zero to one fifth of the ground state energy (of the isolated coupled transmon qutrits) of $\gamma$, the quantum resources increase almost linearly with the increase of $\gamma$. Beyond this weak-coupling (between the subsystems) regime, the rate of increase of maximum self generated quantum resources with increase in \(\gamma\) starts to slow down and approaches saturation to a finite value,
for both the ideal coupled transmon setup isolated from the environments as well as in presence of Markovian environments. In our depiction we have not studied that region of $\gamma$, as the treatment used here is exclusively suitable only for weak coupling regime of the system. In Figs.~\ref{fig:my_label} and~\ref{cc_g} we have shown the cases for the initial state $\psi_s(0)=\ket{0}$, but we have also studied for some more paradigmatic initial separable states. The results are qualitatively similar.




\begin{figure}[h]
\begin{framed}
    
\hspace{-1.5 em}
\includegraphics[width=1\textwidth]{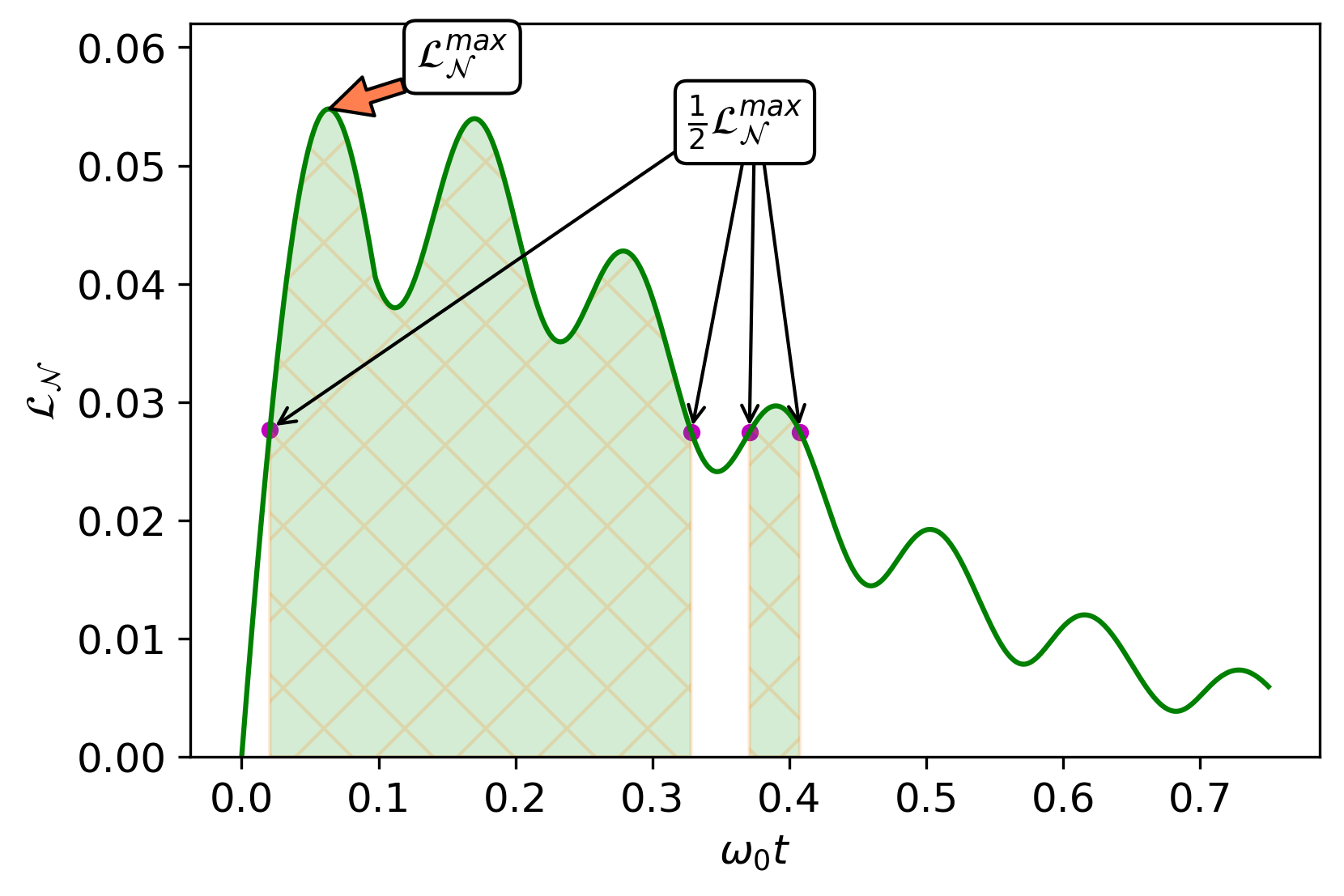} 
\hfill
    \caption{Maximum entanglement-generating power of coupled transmons. Here we plot the logarithmic negativity of entanglement against time for the optimal initial state $\rho_{s_{\text{opt}}}^0$. The optimal state is found to be $\rho_{s_{\text{opt}}}^0=\ket{\psi^1_0} \bra{\psi^1_0} \otimes \ket{\psi^2_0} \bra{\psi^2_0}$ with $\ket{\psi^1_0}= (-0.256-0.492j)\ket{0}+(-0.479+0.680j) \ket{1}$ and $\ket{\psi^2_0}= (-0.395+0.392j)\ket{0}  +(-0.768 -0.316j]) \ket{1}$. 
    All other considerations are same as in Figs.~\ref{fig:1} and~\ref{rr_bb_u}. The quantity plotted along the vertical axis is in ebits and the same plotted against the horizontal axis is dimensionless.}
\label{entopt_1}
\end{framed}
\end{figure}

\section{Optimal entanglement generation with unentangled initial states}
\label{sec:5}

In the previous sections, we have witnessed that two interacting transmon qutrits can generate quantum resources even in presence of decoherence effects of the environment and can sustain a substantial amount of the same for a finite duration of time when starting from 
some zero resource states. 
We have also described that if we start from the states of the system with non-zero resources, they are incapable of generating quantum resources beyond their initial values, which has been verified for a set of entangled paradigmatic initial states.
This leads us to the conclusion that the self generating power of quantum resources of such a system is revealed when the system starts out from zero quantum resource.
In connection to this observation, here we define the entanglement generation power of coupled transmon systems with initial separable states. 

The definition depends on the two aspects: (a) the maximum value of the entanglement the system can self-generate, and (b) the time-scale through which the system can retain a substantial value of this entanglement. Let the maximum generated logarithmic negativity for a separable initial three-qutrit state is $\mathcal{L}_N^{\text{max}}$. We set the threshold to half of the maximum generated  entanglement $\frac{1}{2}\mathcal{L}_N^{\text{max}}$, beyond which the entanglement is assumed to be too small to be useful and hence not taken into account. Therefore, the entanglement generation power of a coupled transmon system as a function of its initial states is given by

\begin{equation}
    \mathcal{E}(\rho_s^0)= \sum_n \int_{t_{n-}}^{t_{n+}}  \mathcal{L_N}(\rho_s^0)dt.
\end{equation}

Here $t_{n^-}$ to $t_{n^+}$ is the time interval during which the system can generate entanglement $\ge \mathcal{L}_{\mathcal{N}}^{\max}/2$ under the \(n^{th}\) peak, and we are taking the summation over all such time intervals to measure entanglement generating power of the system .
In search of the initial state to provide the maximum generation of bipartite entanglement in the system described, we have optimized the quantity and define the optimal entanglement-generating power of the system as
\begin{equation}
    \mathcal{E}_{opt}=\max_{\rho_s^0\in \mathcal{S}} \mathcal{E}(\rho_s^0), 
\label{opteq}
\end{equation}
where $\mathcal{S}$ being the set of all pure product and separable states in $\mathbb{C}^3\otimes \mathbb{C}^3$ Hilbert space.

In Fig.~\ref{entopt_1} we demonstrate the logarithmic negativity of entanglement for the optimal initial state $\rho_{s_{opt}}^0$. The form of $\rho_{s_{opt}}^0$ is described in the caption of the same figure. For this optimal state we get the optimal entanglement-generating power $\mathcal{E}_{opt} \simeq $ 0.01433 ebits. Here we have done optimization over the initial states by choosing $2 \times 10^3$ separable (both pure and mixed states are taken into account separately) states Haar uniformly from the $\mathbb{C}^3\otimes\mathbb{C}^3$ Hilbert space.

\vspace{2em }
\section{Conclusion}
\label{sec:6}

We have studied a system of two transmons, coupled through their charge degrees of freedom.
We have taken into account upto the second excited state of the individual subsystems, appealing to the very small value of charging energy of Cooper pairs, compared to the large Josephson energy, omitting higher energy levels.
The obtained results reported in this work is drawn upon the assumption that the transmons are immersed in independent bosonic baths, within the Markovian limit.

It turns out that such a system can attain a substantial value of entanglement and quantum coherence, even when it starts from states with zero resource.
In the weakly interacting limit, we have also found that for interaction strengths that are small compared to other energy-scales in the system, this self-generated resource in the system increases almost linearly with increasing coupling strength and the long-time value of entanglement and global quantum coherence of the system maintains a significant value. Without the non-local interaction term between the two transmons, entanglement in the system would only decay due to loss of information to the Markovian baths.
However, the interaction present in between the subsystems opposes this decay by feeding resource into the system. We find that even for relatively small non-zero interactions in the system, the entanglement of the composite system does not entirely vanish. 
The competition between the pumping of the resource by the non-local interaction and the loss of the same due to the interaction with the baths results in a sustenance of a stable value of entanglement between the transmons. We also define the optimal entanglement-generating power for a coupled transmon system and obtain the best two-qutrit input state, initialized in the corresponding qubit subspaces,  which can offer the optimal entanglement-generating power of the system under consideration.

\begin{acknowledgments}
 
This research of TR and AG was supported in part by the `INFOSYS scholarship for senior students'. We also acknowledge partial support from the Department of Science and Technology, Government of India through the QuEST  grant (grant number DST/ICPS/QUST/Theme-3/2019/120).

\end{acknowledgments}


\end{document}